\hfuzz 2pt
\font\titlefont=cmbx10 scaled\magstep1
\magnification=\magstep1

\null
\vskip 1.5cm
\centerline{\titlefont NON-STANDARD NEUTRAL KAONS DYNAMICS}
\medskip
\centerline{\titlefont FROM D-BRANE STATISTICS}
\vskip 2.5cm
\centerline{\bf F. Benatti}
\smallskip
\centerline{Dipartimento di Fisica Teorica, Universit\`a di Trieste}
\centerline{Strada Costiera 11, 34014 Trieste, Italy}
\centerline{and}
\centerline{Istituto Nazionale di Fisica Nucleare, Sezione di 
Trieste}
\vskip 1cm
\centerline{\bf R. Floreanini}
\smallskip
\centerline{Istituto Nazionale di Fisica Nucleare, Sezione di 
Trieste}
\centerline{Dipartimento di Fisica Teorica, Universit\`a di Trieste}
\centerline{Strada Costiera 11, 34014 Trieste, Italy}
\vskip 2cm
\centerline{\bf Abstract}
\smallskip
\midinsert
\narrower\narrower\noindent
The neutral kaon system can be effectively described by non-unitary, 
dissipative, completely positive dynamics that extend the usual treatment. 
In the framework of open quantum systems, we show how the origin of these 
non-standard time evolutions can be traced to the interaction of the kaon 
system with a large environment. We find that D-branes, effectively described 
by a heat-bath of quanta obeying infinite statistics, could constitute a 
realistic example of such an environment.
\endinsert
\bigskip
\vfil\eject

{\bf 1. INTRODUCTION}
\medskip

Recently, it has been proposed to describe the system of neutral kaons
using effective dynamics that differ from 
the standard Weisskopf-Wigner one.[1, 2]
These generalized time-evolutions produce loss of quantum coherence and
lead to $CP$ and $CPT$ violating effects that can be in principle 
detected inthe next generations of neutral kaon experiments.

The original physical motivation underlying such approaches come from quantum
gravity that predicts incoherent quantum phenomena at Planck's length
due to the fluctuation of the gravitational field.[3-6] Actually,
such non-standard dynamics could also be the result
of much more general considerations, based on the observation that unstable
systems can be viewed as specific examples of open quantum systems.[7-10]

These systems can be modeled as being small subsystems in ``weak'' interaction
with large environments. Under mild physical assumptions, the reduced dynamics
of the subsystem, obtained by eliminating the environment degrees of freedom,
is realized by dissipative time-evolution maps, with forward in time 
composition law (semigroup property) and 
the additional characteristic of being completely positive. 
This set of transformations forms a so-called dynamicalsemigroup.[11-13]

In the case of the neutral kaon system, whose states 
can be conveniently described
by $2\times 2$ density matrices $\rho$,[14] the above general considerations
lead to the following effective time-evolution equation:[7]
$$
{\partial\rho(t)\over \partial t}= -i \big(H_{\rm eff}\ \rho(t)-\rho(t)\,
H_{\rm eff}^\dagger\big) + D[\rho(t)]\ .\eqno(1.1)
$$
The effective hamiltonian $H_{\rm eff}$ (the Weisskopf-Wigner hamiltonian)
accounts for the kaon decays and includes a non-hermitian part, that
characterizes the natural width of the kaon states.
The non-standard piece $D[\rho]$ is a linear, trace-preserving map, 
transforming density matrices into density matrices. 
It has the general form:
$$
D[\rho]={1\over2} \sum_{i,j=1}^3 a_{ij}\big[2\, \sigma_j\rho\,\sigma_i 
-\sigma_i\sigma_j\, \rho -\rho\,\sigma_i\sigma_j\big]\ ,\eqno(1.2)
$$
where $\sigma_i$, $i=1,2,3$, are the Pauli matrices and the coefficients
$a_{ij}$ form a real, symmetric, positive definite matrix $[a_{ij}]$.
They can be expressed in terms of six phenomenological parameters,
obeying certain inequalities.[7] The magnitude of these parameters is
surely very small: the contribution $D[\rho]$ in (1.1) has yet to be
directly detected and therefore it should be considered as subleading 
with respect to the standard $H_{\rm eff}$ term.

The positive definiteness of $[a]$ guarantees the complete positivity of the
map $D[\rho]$, and hence of the time-evolution produced by (1.1); 
this property assures not only the positivity for all times of the 
eigenvalues of $\rho(t)$,
but also of any density matrix describing systems of correlated kaons.
On the other hand, the reality of the coefficients 
$a_{ij}$ assures the increase
in time of the von Neumann entropy: 
$-d/dt \, {\rm Tr}[\rho(t)\ln\rho(t)]\geq 0$.[8]

The phenomenological consequences of the 
time-evolution (1.1) have been studied 
in detail, in particular in view of their possible experimental measure
at $\phi$-factories.\hbox{[9, 10]} 
In the following, we shall devote our attention to the
discussion about the origin of the additional dissipative 
term $D[\rho]$ in (1.1).
As mentioned before, this term can be considered as 
the consequence of the incoherent
interaction of the kaon system with an environment. In the next sections,
we shall study in detail all the steps and procedures 
that justify this result.
The outcome of this analysis is that a heat-bath of D-branes obeying the 
so-called infinite statistics could be the ultimate 
origin of the non-standard
contribution (1.2). The experimental study of the effects of this term could
therefore provide important (although indirect) informations on the
underlying dynamics of low energy string theory.

\vskip 2cm

{\bf 2. THE MASTER EQUATION}
\medskip

Let us consider a system $S$, later to be identified with the neutral
kaon system, in interaction with a large environment $E$. At this stage,
we shall try to be as general as possible, and therefore we leave
the basic characteristics of $E$ unspecified. The hamiltonian $H$ of the
global system $S+E$ can be in general decomposed as
$$
H=H_S\otimes {\bf 1} + {\bf 1}\otimes H_E + H'\ ,\eqno(2.1)
$$
where $H_S$ describes the dynamics of $S$ in absence of the environment,
while $H_E$ that of $E$; the term $H'$ takes care of the interaction between
the two systems, that is assumed to be weak.

The system $S$ could be unstable, as indeed is the case for the kaons,
and therefore the hamiltonian $H_S$ needs not be hermitian. On the contrary,
the contribution of the environment to possible 
decay processes can be estimated
to be negligible for any practical considerations;[15] specifically, the
kaon decay is driven by the weak interactions and not by the effects of the
environment. The description of the kaon self-dynamics $H_S$ by means
of a Weisskopf-Wigner hamiltonian is therefore appropriate and will be
adopted in the following. On the other hand, there is no restriction in taking
the environment hamiltonian $H_E$ to be hermitian.

Moreover, since the kaons are produced via the
strong interactions, it is natural to further assume that $S$ and $E$ be
uncorrelated at the moment of the formation of the unstable system $S$.
This implies that initially the density matrix $\rho_{S+E}$ describing the
state of the global system $S+E$ is in factorized form: 
$\rho_{S+E}=\rho_S\otimes\rho_E$.

The time evolution of the total system $S+E$ is simply given by the equation:
$$
{\partial\rho_{S+E}(t)\over\partial t}=-i\big( H\, \rho_{S+E}(t)
-\rho_{S+E}(t)\, H^\dagger\big)\ ,
\eqno(2.2)
$$
or, in finite form,
$$
\rho_{S+E}(t)=e^{-iHt}\ \big(\rho_s\otimes\rho_E\big)\ e^{iH^\dagger t}\ .
\eqno(2.3)
$$
However, we are interested in the dynamics of the subsystem $S$ only; this can
be obtained by tracing over the environment degrees of freedom:
$\rho_S\rightarrow\rho(t)\equiv {\rm Tr}_E\Big(\rho_{S+E}(t)\Big)$.
To obtain the equation that describes this reduced time-evolution,
it is convenient to introduce a vector notation, rewriting the density matrix
$\rho_{S+E}$ for the global system as the vector $|\rho_{S+E}\rangle$.
The operation of tracing over the environment degrees of freedom can now be
represented by a projector operator $P$, so that the state vector of the
subsystem $S$ is simply:
$$
|\rho\rangle\equiv P\, |\rho_{S+E}\rangle=
\big|{\rm Tr}_E\big(\rho_{S+E})\big\rangle\ ,\eqno(2.4)
$$
with $P^2=P$; the complement projector $1-P$ will be denoted by $Q$. 
By introducing the generalized Liouville operator $L_H$, 
corresponding to the total hamiltonian $H$,
$$
L_H\ |\rho_{S+E}\rangle\equiv\big| H\, \rho_{S+E}-
\rho_{S+E}\, H^\dagger\big\rangle\ ,\eqno(2.5)
$$
one can write (2.2) as a Schr\"odinger-like equation:
$$
i{\partial\over\partial t}|\rho_{S+E}(t)\rangle=L_H\ |\rho_{S+E}(t)\rangle\ .
\eqno(2.6)
$$
Notice that one can identify $P|\rho_{S+E}(0)\rangle$ again with 
$|\rho_{S+E}(0)\rangle$, since at $t=\,0$ the state of $S+E$ is factorized;
as a consequence, one has: $Q|\rho_{S+E}(0)\rangle=\,0$. All these steps
can be rigorously justified from the mathematical point of view; however,
for sake of simplicity we shall keep all technical discussions to a minimum.

The equation describing the subdynamics for $S$ can now be easily obtained
by projecting (2.6) on $P$ and $Q$. Using the Laplace transformed vector
$|\widetilde\rho\,\rangle$ of (2.4), one finds:
$$
\bigg[z+i\bigg(L_{PP}+L_{PQ}\ {1\over iz-L_{QQ}}\ L_{QP}\bigg)\bigg]\ 
\big|\widetilde\rho(z)\big\rangle=\big|\widetilde\rho(0)\big\rangle
\equiv\big|\widetilde\rho_S\big\rangle\ ,\eqno(2.7)
$$
where $z$ is the Laplace variable and
$$
L_{PP}=PL_H P\ ,\qquad L_{QP}=QL_H P\ ,\qquad L_{PQ}=PL_H Q\ ,\qquad
L_{QQ}=QL_H Q\ .\eqno(2.8)
$$
From this result, using the inverse transform, 
one immediately obtains the ``master equation'' describing
the time evolution of the projected vector $|\rho(t)\rangle$, representing
the state of the subsystem $S$:
$$
i{\partial\over\partial t}|\rho(t)\rangle=L_{PP}\ |\rho(t)\rangle 
-i\int_0^t dt'\ L_{PQ}\, e^{-i\,L_{QQ}\, (t-t')}\, L_{QP}\ |\rho(t')\rangle\ .
\eqno(2.9)
$$

One can prove that the evolution map $\rho_S\rightarrow\rho(t)$ given by 
this equation is completely positive,[11-13] 
as the one produced by the equation (1.1).
However, (2.9) is clearly not of the form (1.1): although the term
$L_{PP}\, |\rho(t)\rangle\equiv \big(L_{H_S}\big)_{PP}\, |\rho(t)\rangle$ 
(see below) can be identified
with the Weisskopf-Wigner part of (1.1), the second term in the r.h.s. of
(2.9) is not of Markovian form; in fact, $|\rho(t)\rangle$ as given by (2.9)
depends not only on the initial state $|\rho(0)\rangle$, but also on all
the states $|\rho(t')\rangle$, with $t'<t$. This is a general result of any
reduced dynamics: the form of the master equation (2.9) 
can be further simplified
only on the basis of additional physical considerations. In the present case,
these additional assumptions correspond to the requirement 
that the interaction
between the system $S$ and the environment $E$ be weak. 
Indeed, as we shall see
in the next section, this condition allows the suppression of all
memory effects.

\vfill\eject

{\bf 3. THE WEAK COUPLING LIMIT}
\medskip

In writing down the hamiltonian (2.1) for the total system $S+E$ we have
assumed the interaction between $S$ and $E$ to be weak. There are essentially
two different ways of implementing this physical requirement in the general
master equation (2.9).[11-13] In the first approach, 
one multiplies the interaction
hamiltonian $H'$ by a small dimensionless coupling constant $\lambda$; 
in this way,
one treats the motion of the system $S$ of order one, while the dissipative
part of (2.9) (the second term in the r.h.s.) 
turns out to by of order $\lambda^2$.
In view of this, the evolution of the state $|\rho(t)\rangle$ has 
to be studied
on a time scale $1/\lambda^2$. In order to consistently 
implement this condition,
one rescales the time variable, $t\rightarrow t/\lambda^2$, and then takes
the limit $\lambda\rightarrow 0$. The idea is to use these steps to extend
the integral in (2.9) to infinity and to eliminate the dependence on $t'$.
This procedure is known as Markov approximation; 
it requires some care because
in general it is not uniquely defined.[16]

For a more complete discussion, it is useful to define two time scales:
$\tau_S$, the typical variation time of $|\rho(t)\rangle$ given by the
``free'' evolution $L_{PP}$ in (2.9), and $\tau_E$, the decay time of the
correlations in the environment $E$. On general grounds, one expects
the memory effects in (2.9) to be negligible if $\tau_S$ is much longer
than $\tau_E$, or more precisely when the ratio $\tau_S/\tau_E$ becomes
very large. And indeed, the limit sketched above corresponds to sending
$\tau_S$ to infinity, while keeping $\tau_E$ finite.

In the case of an unstable subsystem $S$, like the neutral kaon system,
$\tau_S$ can be identified with the corresponding lifetime, which is
necessarily finite. The limiting procedure that corresponds to letting
$\tau_S\rightarrow\infty$ is clearly inappropriate and can not be considered
in these situations. 

However, in order to attain $\tau_S/\tau_E\rightarrow\infty$, one can also
let $\tau_E$ to become small, while leaving $\tau_S$ finite. This situation
corresponds to the second weak coupling limit approach, which is obtained
by assuming that the uncoupled motion of the system $S$ and of the dissipative
terms produced by the interaction with the environment are of the same order
of magnitude. As we shall see in the next sections, this limiting procedure
is realized when the typical time-correlations of the environment approach
a $\delta$-function; this explains why in the literature 
this kind of weak coupling
limit is also referred to as ``singular coupling limit''.

Introducing again a dimensionless coupling constant $\lambda$, the total 
hamiltonian can now be written as:[13]
$$
H=\lambda^2\, H_S\otimes {\bf 1} + {\bf 1}\otimes H_E + \lambda\, H'\ ,
\eqno(3.1)
$$
and the Liouville operator $L_H$ appearing in (2.6) can be correspondingly
decomposed as: $L_H=\lambda^2\, L_{H_S} + L_{H_E} + \lambda\, L_{H'}$.
In this case, it is convenient to pass to a rescaled time variable,
$t\rightarrow t/\lambda^2$, directly in the original equation (2.6),
so that the master equation (2.9) becomes:
$$
i{\partial\over\partial t}|\rho(t)\rangle=
{1\over\lambda^2}L_{PP}\ |\rho(t)\rangle 
-{i\over\lambda^4}\int_0^t dt'\  L_{PQ}\, e^{-i\, L_{QQ}\, t'/\lambda^2}\, 
L_{QP}\ |\rho(t-t')\rangle\ .
\eqno(3.2)
$$
By letting $t'=\lambda^2\tau$ and taking
the limit $\lambda\rightarrow0$, the previous equation reduces to:
$$
i{\partial\over\partial t}|\rho(t)\rangle=\bigg[\big(L_{H_S}\big)_{PP} 
-i\int_0^\infty d\tau\ \big(L_{H'}\big)_{PQ}\, e^{-i(L_{H_E})_{QQ}\,\tau}\, 
\big(L_{H'}\big)_{QP}\bigg]\ |\rho(t)\rangle\ ,
\eqno(3.3)
$$
where the conditions $L_{PP}=\big(L_{H_S}\big)_{PP}$ and
$L_{PQ}=\big(L_{H'}\big)_{PQ}$, $L_{QP}=\big(L_{H'}\big)_{QP}$
have been used. These properties follow from 
$$
P\, L_{H'}\, P=\,0\ ,\qquad P\, L_{H_E}=L_{H_E}\, P=\,0\ ,\qquad
P\, L_{H_S}=L_{H_S}\, P\ .\eqno(3.4)
$$
The first relation is equivalent to the assumption that the interaction
hamiltonian $H'$ has no diagonal elements in the representation in which
$H_E$ is diagonal; in practice, this is not really a restriction, 
since we can
always redefine $H_S$ and $H'$ in such a way to absorb such diagonal terms
in $H_S$. The second relation in (3.4) is a 
consequence of probability conservation
in the environment (first half) and of the condition of thermal equilibrium,
$L_{H_E} |\rho_E\rangle\equiv|[H_E,\,\rho_E]\rangle=\,0$, 
to be discussed in the next section (second half). Finally, 
the third condition above
holds since $P$ and $L_{H_S}$ act in different spaces.

The equation  (3.3) can be further simplified by using 
$Q\, L_{H_E}\, Q= L_{H_E}$
which is a consequence of (3.4).
Coming back to the standard density matrix notation, one finally finds:
$$
{\partial\rho(t)\over \partial t}= -i \big(H_S\ \rho(t)-\rho(t)\, 
H_S^\dagger\big) + D[\rho(t)]\ ,\eqno(3.5)
$$
with
$$
D[\rho(t)]=-\int_0^\infty d\tau\ {\rm Tr}_E\bigg\{\Big[ e^{iH_E\tau}\, H'\, 
e^{-iH_E\tau}\ , \big[H',\rho(t)\otimes\rho_E\big]\Big]\bigg\}\ .
\eqno(3.6)
$$
This evolution equation is of markovian form: all memory effects have
disappeared. The dissipative term $D[\rho]$ involve an integration
over environment ``time-correlations'' of the interacting hamiltonian $H'$.
Its explicit form will depend on the structure of $H'$ and 
of the characteristic
properties of the environment. In the next sections we shall specialize
the rather general evolution equation (3.5), (3.6) to the case of the
neutral kaons.

\vskip 2cm

{\bf 4. THE ROLE OF THE ENVIRONMENT}
\medskip

As it was mentioned in the introductory remarks, 
in the case of elementary particle
systems it is natural to associate the effects of 
the environment $E$ as being
gravitational in origin. Despite the smallness of the gravitational coupling,
the quantum fluctuations at Planck's length, {\it e.g.} realized via the 
space-time foam, could indeed act as a weak coupled environment.[17]

From a more fundamental viewpoint, such gravitational effects are likely to
originate from the dynamics of extended objects, strings or branes, and could
in principle be deduced using low energy string theory. Although attempts
in this direction have been presented in the literature,[18] 
for our considerations it will be
sufficient to have an ``effective'' description of such an environment,
taking into account the most fundamental characterizing properties of 
the underlying string dynamics.

To be more specific, we shall model the environment as a heat-bath, 
{\it i.e.}
a gas of non-interacting bosonic quantum modes, 
in thermodynamical equilibrium
at Planck's temperature $\beta_P^{-1}\sim M_P$, the natural scale at which
the dissipative effects can be assumed to start being relevant.
The environment initial state is then proportional to the Gibbs distribution:
$$
\rho_E={e^{-\beta_P H_E}\over {\rm Tr}_E\big(e^{-\beta_P H_E}\big)}\ .
\eqno(4.1)
$$
Moreover, by assumption, the environment is 
very large so that its statistical 
properties are essentially unaffected 
by the weak coupling to the system $S$.%
\footnote{$^\dagger$}{Actually, in order to provide a sensible
``singular coupling limit'' the thermal bath must be infinite, in which
case the state (4.1) does not make sense; however, all calculations performed
with (4.1) can be properly generalized also to the case of the infinite bath.
[19-22]}
In other terms, the environment remains in equilibrium and its state is given
by (4.1) also at later times. Notice that this implies $[H_E,\, \rho_E]=\,0$,
as used in checking the second relation in (3.4).

The form of the hamiltonian $H'$ that describes the interaction of $S$, 
henceforth identified with the neutral kaon system, with such an environment
is largely arbitrary. The only physical requirement that needs to be taken into
account is that $H'$, when inserted in (3.6), must produce ``small'' effects 
on the system $S$, which is for the most part driven by the effective
Weisskopf-Wigner hamiltonian term in (3.5).
In view of these considerations, 
it is natural to choose an hamiltonian $H'$ that is
linear in the kaon and environment observables.

As already observed, the kaon system can be effectively represented by
means of a two-dimensional Hilbert space.[14] 
With a suitable choice of basis vectors in
this space, the interaction $H'$ can be written in the following generic form:
$$
H'=g\, \sum_{\mu=0}^3 \sigma_\mu\otimes B_\mu\ ,\eqno(4.2)
$$
where as before $\sigma_i$, $i=1,2,3$, are the Pauli matrices, 
with $\sigma_0$ the
$2\times2$ unit matrix, and $B_\mu$ is an hermitian operator describing
the environment, whose explicit expression will be discussed later;
$g$ is a dimensionless coupling constant, that should be expressible in terms
of the relevant mass scales, {\it i.e.} the kaon mass $m_K$ 
and the Planck mass
$M_P$. Since $g$ is small, it must be of order $(m_K/M_P)^\delta$, with
$\delta$ a fixed power; as a working assumption, in the following 
we shall take $\delta\sim 1$.

At this point, one can insert (4.2) in (3.6) and simplify the
expression of the dissipative term $D[\rho]$. Since the hamiltonian $H_E$
acts only on the environment degrees of freedom, one first observes that:
$$
e^{iH_E t}\ H'\ e^{-iH_E t}=g\, \sum_{\mu=0}^3 \sigma_\mu\otimes B_\mu(t)\ .
\eqno(4.3)
$$
Then, expanding the double commutator in (3.6) 
and performing the trace operation
over the environment degrees of freedom, one finally obtains:
$$
\eqalign{
D[\rho]=g^2\, \sum_{\mu,\nu} \int_0^\infty dt 
\Big[ &-\sigma_\mu\sigma_\nu\,\rho\ 
\langle B_\mu(t)\, B_\nu\rangle + \sigma_\mu\, \rho\, \sigma_\nu\ 
\langle B_\nu\, B_\mu(t)\rangle \cr
&+ \sigma_\nu\, \rho\, \sigma_\mu\
\langle B_\mu(t)\, B_\nu\rangle - \rho\, \sigma_\nu\sigma_\mu\
\langle B_\nu\, B_\mu(t)\rangle\Big]\ ,}\eqno(4.4)
$$
where
$$
\langle B_\mu(t)\, B_\nu\rangle={\rm Tr}_E\Big[B_\mu(t)\,
B_\nu\,\rho_E\Big]\ ,
\eqno(4.5)
$$
are the time-correlation functions of the environment operators.
Using the definition (4.5), one can easily verify 
that the following properties hold:
$$
\langle B_\mu(t)\, B_\nu\rangle^*=\langle B_\nu\, B_\mu(t)\rangle=
\langle B_\nu(-t)\, B_\mu\rangle\ .\eqno(4.6)
$$
Further, by introducing the following $4\times4$ hermitian matrices, 
with entries:
$$
\eqalignno{
&\alpha_{\mu\nu}=\int_0^\infty dt\ \langle B_\mu(t)\, B_\nu\rangle
+ \int_0^\infty dt\ \langle B_\mu\, B_\nu(t)\rangle\equiv
\int_{-\infty}^\infty dt\ \langle B_\mu(t)\, B_\nu\rangle\ , &(4.7a)\cr
&\beta_{\mu\nu}=i\bigg(\int_0^\infty dt\ \langle B_\mu(t)\, B_\nu\rangle
- \int_0^\infty dt\ \langle B_\mu\, B_\nu(t)\rangle\bigg)\ , &(4.7b)}
$$
one can rewrite (4.4) as:
$$
D[\rho]={g^2\over2}\,\sum_{i,j=1}^3 \alpha_{ij}\, 
\big[2\, \sigma_j\rho\, \sigma_i 
-\sigma_i\sigma_j\,\rho -\rho\,\sigma_i\sigma_j\big]
+{ig^2\over2}\big[\rho\ , \widetilde H\big]\ ,
\eqno(4.8)
$$
where the hermitian operator $\widetilde H$ is 
explicitly given by the combination:
$$
\widetilde H=\sum_{\mu,\nu=0}^3\beta_{\mu\nu}\, \sigma_\mu\sigma_\nu
+i\sum_{i=1}^3\big(\alpha_{0i}-\alpha_{i0}\big)\,\sigma_i\ .
\eqno(4.9)
$$
The second term in (4.8) is hamiltonian in character and can be reabsorbed in
a redefinition of the hermitian part of the kaon effective hamiltonian
$H_S$ in (3.5). The remaining piece is the true dissipative term. The
equation (3.5) that describes the time-evolution of the kaon
$2\times2$ density matrix $\rho(t)$ has now the form (1.1),
(1.2) presented in the Introduction, although the precise structure
of the coefficients $\alpha_{\mu\nu}$ in $(4.7a)$ has yet to be addressed.

We would like to stress that this evolution equation
has been derived by means of standard techniques used in the analysis
of open quantum systems,[11-13] more specifically in quantum optics:[23, 24] 
with the help of rather general assumptions, 
motivated by plausible physical arguments, 
we have been able to show that these 
universal tools can be effectively applied 
to describe also elementary particle systems, like the neutral kaons.
As we shall see in the next section, this description could be
motivated more precisely by certain general aspects of the 
low energy dynamics of string theory.

\vfill\break

{\bf 5. D-BRANES AS ENVIRONMENT}
\medskip

The structure of the integrated correlation functions $\alpha_{\mu\nu}$
in $(4.7a)$ are essentially determined by the statistics obeyed by the quanta
that constitute the heat-bath, which effectively describe
the environment. In $n$ space-time dimensions, the bath operator $B_\mu$,
that as explained before should be taken to be linear in the modes
creation and annihilation operators, can be assumed to have the following
very general structure:
$$
B_\mu(t)={1\over M_P^{(n-4)/2}}\sum_a
\int {d^{n-1}k\over [2(2\pi)^{n-1}\omega(k)]^{1/2}}
\, f^a(k)\, \Big[\chi_\mu^a\ A_a(k)\, e^{-i\omega(k) t}
+(\chi_\mu^a)^*\ A_a^\dagger(k)\, e^{i\omega(k) t}\Big].\eqno(5.1)
$$
The coefficients $\chi_\mu^a$ ``embed'' the bath modes 
into the two-dimensional
kaon Hilbert space and can be taken to satisfy
$$
\sum_a (\chi_\mu^a)^*\ \chi_\nu^a=\delta_{\mu\nu}\ ,\eqno(5.2)
$$
while $f^a(k)$ are appropriate test functions necessary to make the
operator $B_\mu$ and its correlations well-defined; 
in general, it can be taken to be of the form $(|k|/M_P)^{m/2}\, g^a(k)$,
for some positive integer $m$, with $g^a(k)$ of Gaussian type.
The function $\omega(k)$ gives the dispersion relation
obeyed by the bath modes; in the following, for simplicity we shall take 
an ultrarelativistic law: $\omega(k)=|k|\equiv\omega$.%
\footnote{$^\dagger$}{Possible infrared infinities can be 
avoided by a suitable
choice of the integer $m$ in the test function $f^a(k)$.}
The powers of Planck's
mass, characterizing the energy scale of the bath, are necessary to give
$B_\mu$ the right dimension of energy.

As implicit in the form of the time evolution in (5.1), 
the creation $A_a^\dagger$
and annihilation $A_a$ operators for the bath modes 
fulfill the following general
commutation relations with the environment hamiltonian $H_E$:
$$
[H_E, A_a^\dagger(k)]=\omega(k)\, A_a^\dagger(k)\ ,\qquad
[H_E, A_a(k)]=-\omega(k)\, A_a(k)\ .\eqno(5.3)
$$
However, from these relations one can not infer any conclusion concerning
the nature of the quanta in the bath, 
and in particular about their statistics.
Indeed, it has been observed long ago that statistics other than that of Bose
and Fermi could be envisaged within standard many-body quantum mechanics;[25]
this fact has led to the study of the so 
called ``parastatistics'' (see [26, 27]
and references therein).
The modes of the elementary excitations obeying these generalized statistics
are quantized using commutation relations that are trilinear in the creation
and annihilation operators, and whose representations are characterized by
an integer $p$, the so called order of parastatistics. This number basically 
corresponds to the number of quanta in a given 
symmetric or antisymmetric state.

A further generalization occurs in the case of 
the ``infinite statistics'',[28]
that correspond to the case where the number $p$ is left as a free parameter.
They are realized by $q$-oscillator algebras, 
{\it i.e.} by one-parameter extensions
of the standard oscillator algebra.

These generalized infinite statistics 
have recently attracted a lot of attention
in string theory and in particular in $M$(atrix) theory. 
This interest originates
from the observation that the thermodynamical properties of black holes
can be obtained studying the statistics of effective string models, described
in terms of D-branes.[29-33, 34-37] It has been recognized 
that these extended objects
(and in particular D0-branes, the quanta of Matrix theory) satisfy quantum
infinite statistics,[38, 39, 36] and this result further clarify 
the D-brane interpretation
of neutral black holes thermodynamics.

More in general, D-branes are shown to capture in a non-perturbative way
many low energy properties of $M$-theory,[40] 
and therefore should provide the
right description for studying quantum gravity effects at Planck's scale.

In view of these considerations, it seems plausible 
to assume that the heat-bath
responsible for the appearance of the dissipative term (3.6) in the effective
evolution equation for the $K^0$-$\overline{K^0}$ system is actually given
by an ensemble of D-branes; in this case, the creation and annihilation
operators in (5.1) should obey an infinite statistics. As previously noted,
this situation can be realized by a $q$-oscillator algebra, that in our case
can be presented in the following general form:[41]
$$
A_a(k)\, A_b^\dagger(k')- q\, A_b^\dagger(k')\, A_a(k)=
\delta_{ab}\ \delta^{(n-1)}(k-k')\ ,\eqno(5.4)
$$
where $q$ is the deformation parameter; to assure the reality of the
operator $B_\mu$ in (5.1), $q$ must be real, and without loss of generality
we can take $q\leq 1$. The case $q=1$ corresponds to standard bosons,
while for $q=0$ one obtains the degenerate algebra discussed in [38, 39, 36],
in connection with D0-branes and black holes.
In the computations that follows we shall therefore assume $q<1$.
The single-mode hamiltonian is taken to be proportional 
to the corresponding number operator, 
so that the total hamiltonian $H_E$ indeed satisfies the relations (5.3).

We are interested in computing the correlation functions 
$\langle B_\mu(t) B_\nu\rangle$; using (5.1), 
these can be expressed in terms
the $A^\dagger$-$A$ thermal correlations. A straightforward computation 
shows that:[41, 42]
$$
\langle A_a^\dagger(k)\ A_b(k')\rangle=\delta_{ab}\ \delta^{(n-1)}(k-k')\ 
N_q\big(\omega(k)\big)\ ,\eqno(5.5)
$$
where
$$
N_q\big(\omega(k)\big)= {1\over e^{\beta_P\omega(k)}-q}\ ,\eqno(5.6)
$$
is the generalized ensemble distribution. Explicit use of (5.4) further gives:
$$
\langle A_a(k)\ A_b^\dagger(k')\rangle=\delta_{ab}\ \delta^{(n-1)}(k-k')\ 
e^{\beta_P\omega(k)}\ N_q\big(\omega(k)\big)\ ,\eqno(5.7)
$$
while $\langle A_a A_b\rangle$ and $\langle A_a^\dagger A_b^\dagger\rangle$
vanish, as follows from the orthogonality of states with different
occupation number.

Using these relations and introducing 
$(n-1)$-dimensional spherical coordinates,
with the help of (5.1) one obtains:
$$
\big\langle B_\mu(t)\ B_\nu\big\rangle=
{1\over2\, M_P^{m+n-4}}\int_0^\infty d\omega\
\omega^{m+n-3}\ \Big[X_{\mu\nu}^*(\omega)\, e^{-i\omega(t+i\beta_P)}+
X_{\mu\nu}(\omega)\, e^{i\omega t}\Big]\ N_q(\omega)\ ,\eqno(5.8)
$$
while the integration over the angle variables gives,
$$
X_{\mu\nu}(\omega)=\sum_a\bigg[\int{d\Omega_{n-1}\over(2\pi)^{n-1}}\
\big(g^a(k)\big)^2\bigg]\ (\chi_\mu^a)^*\, \chi_\nu^a
\equiv X_{\nu\mu}^*(\omega)\ .\eqno(5.9)
$$
Notice that the matrix $[X_{\mu\nu}]$ is positive definite. Further,
the expression (5.8) for the environment correlations makes it transparent
that the following periodicity property holds:
$$
\big\langle B_\mu(t)\ B_\nu\big\rangle=
\big\langle B_\nu\ B_\mu(t+i\beta_P)\big\rangle=
\big\langle B_\nu(-t-i\beta_P)\ B_\mu\big\rangle\ .\eqno(5.10)
$$
The first equality amounts to the so-called 
Kubo-Martin-Schwinger (KMS) condition,[43]
that holds for our choice of the D-brane thermal state (4.1), 
while the second
equality just expresses the time-invariance of 
the correlation functions in that
equilibrium state.

The matrix coefficients $\alpha_{ij}$, $i,j=1,2,3$, 
appearing in the dissipative term in (4.8)
can now be written in a more explicit form; in fact, 
from the definition $(4.7a)$ and (5.8) one obtains:
$$
\alpha_{ij}
={1\over 2\, M_P^{m+n-4}}\ \int_{-\infty}^\infty dt\ \int_0^\infty d\omega\
\omega^{m+n-3}\ \cos(\omega t)\
\Big[X_{ij}^*(\omega)\, e^{\beta_P\omega}+
X_{ij}(\omega)\Big]\ N_q(\omega)\ .\eqno(5.11)
$$
This double integral converges provided the function multiplying the cosine 
in the integrand satisfies some mild regularity conditions.
To be specific, these technical conditions requires
$\omega^{m+n-3}\,
\big[X_{ij}^*(\omega)\, e^{\beta_P\omega}+
X_{ij}(\omega)\big]\, N_q(\omega)$ to be integrable and of bounded variation
on the real half-line.[44, 45] One can check that, by suitably fixing
the positive integer $m$ so that $m+n\geq 3$, 
these requirements are satisfied by
our choice of the test function $f^a$ in (5.1). 
In particular, general considerations in 
Fourier analysis allows us to conclude
that $\alpha_{ij}$ in (5.11) is non-vanishing 
only when $m=3-n$, and in this case:
$$
\alpha_{ij}={\pi\over 1-q}\, M_P\, x_{ij}\ ,
\eqno(5.12)
$$
where
$$
x_{ij}={1\over2}\Big(X_{ij}(0) + X_{ij}^*(0)\Big)\ ,\eqno(5.13)
$$
turns out to be a real, symmetric, positive definite matrix. 
From (5.9), it is clear that the structure of this matrix 
depends on the specific choice for
the test function $g^a$ and the coefficients $\chi_\mu^a$. 
A more detailed
analysis of the D-brane dynamics would certainly provide 
more informations
on $x_{ij}$ and we hope to come back to these issues in the future.
However, for the scopes of the present investigation, 
it is sufficient to note that
the matrix $[x_{ij}]$ satisfies the same properties 
required for the matrix $[a_{ij}]$ 
in (1.2), and that its entries are of order one. 

At this point, we can summarize our results by 
writing down the explicit form
that the matrix coefficients $a_{ij}$ in (1.2) take as a consequence of
our considerations.
Indeed, comparing (5.12) and (4.8) with
the expression in (1.2) and recalling that the coupling constant $g$ is of
the order $\sim(m_K/M_P)^\delta$, with $\delta\sim 1$, 
one can now identify the coefficient matrix
$[a_{ij}]$, $i,j=1,2,3$ as:
$$
a_{ij}={\pi\over 1-q}\, g^2\, M_P\, x_{ij}=
{\pi\over 1-q}\, {m_K^2\over M_P}\ \Big({m_K\over M_P}\Big)^{2(\delta-1)}\, 
x_{ij}\ .\eqno(5.14)
$$
From this discussion, we can therefore predict that the coefficients that
determine the magnitude of the dissipative term in the non-standard 
evolution equation (1.1) should be 
at most of the order $m_K^2/M_P\sim 10^{-19}\ {\rm GeV}$.

An interesting outcome of the previous analysis
is that the assumption of a D-brane origin for the dissipative effects
in the kaon dynamics gives a constraint on the dimension 
of the space in which the
quanta of the heat-bath, describing the incoherent
effects of the environment, effectively move. Indeed, as discussed above,
the non-vanishing result (5.12) directly implies the bound $n\leq 3$.
Clearly, this result is tight to the statistical properties of those quanta,
direct consequence of the underlying D-brane dynamics.
By changing the statistics obeyed by the heat-bath modes,
one also changes the constraint on the dimension of the space 
in which those modes live.

For the standard Bose statistics, obtained from (5.4)
by setting $q=1$, the evaluation of the integral in (5.11)
would produce the result, 
$\alpha_{ij}=\pi\, M_P\, x_{ij}$, with the bound $n\leq 4$.
In this case, the dynamics of the D-branes heat-bath would be effectively 
described by scalar ``gravitons'' living in an effective
space-time of at most dimension four. 
Finally, if one chooses to work with a different kind of infinite
statistics, described by the algebra that for a single mode reads, 
$A\, A^\dagger - q\, A^\dagger\, A=q^{-{\cal N}}$
($\cal N$ being the number operator), which is another possible
realization of the $q$-oscillator algebra,
then the constraint $n\leq 2$ would be selected by the evaluation of (5.11).
In other terms, these arguments seem to suggest that
the presence of the extra dissipative term (1.2) in the
kaon evolution equation (1.1) is related to the dimension
of the space where the elementary excitations of the D-brane
heat-bath effectively move and to the statistics they obey.

These results are interesting in view of possible 
experimental measures of the
non-standard term in (1.2). 
As discussed in detail in [9, 10], if the coefficents
$a_{ij}$ are really of order $m_K^2/M_P$, then they should be in the reach
of the next generation of $K$ experiments, in particular those that study
correlated kaons at $\phi$-factories. 
If these phenomenological expectations turn out
to be actually confirmed by the experiment, they will be not only of great
importance for the understanding of the kaon physics, but they could also
provide, as the considerations above seem to indicate, 
informations on the effective dynamics of low energy
string theory. We find this possibility one of the most intruiging results
of our analysis.

\vfill\eject

\centerline{\bf REFERENCES}
\bigskip\medskip

\item{1.} J. Ellis, J.S. Hagelin, D.V. Nanopoulos and M. Srednicki,
Nucl. Phys. {\bf B241} (1984) 381; J. Ellis, J.L. Lopez, N.E. Mavromatos 
and D.V. Nanopoulos, Phys. Rev. D {\bf 53} (1996) 3846
\smallskip
\item{2.} P. Huet and M.E. Peskin, Nucl. Phys. {\bf B434} (1995) 3
\smallskip
\item{3.} S. Hawking, Comm. Math. Phys. {\bf 87} (1983) 395; Phys. Rev. D
{\bf 37} (1988) 904; Phys. Rev. D {\bf 53} (1996) 3099
\smallskip
\item{4.} S. Coleman, Nucl. Phys. {\bf B307} (1988) 867
\smallskip
\item{5.} S.B. Giddings and A. Strominger, Nucl. Phys. {\bf B307} (1988) 854
\smallskip
\item{6.} M. Srednicki, Nucl. Phys. {\bf B410} (1993) 143
\smallskip
\item{7.} F. Benatti and R. Floreanini, Nucl. Phys. {\bf B488} (1997) 335
\smallskip
\item{8.} F. Benatti and R. Floreanini, 
Mod. Phys. Lett. {\bf A12} (1997) 1465;
Complete positivity and the neutral kaon system, 
Banach Center Publications, 1998,
to appear
\smallskip
\item{9.} F. Benatti and R. Floreanini, Phys. Lett. {\bf B401} (1997) 337
\smallskip
\item{10.} F. Benatti and R. Floreanini, Nucl. Phys. {\bf B511} (1998) 550
\smallskip
\item{11.} R. Alicki and K. Lendi, {\it Quantum Dynamical Semigroups and 
Applications}, Lect. Notes Phys. {\bf 286}, (Springer-Verlag, Berlin, 1987)
\smallskip
\item{12.} V. Gorini, A. Frigerio, M. Verri, A. Kossakowski and
E.C.G. Surdarshan, Rep. Math. Phys. {\bf 13} (1978) 149 
\smallskip
\item{13.} H. Spohn, Rev. Mod. Phys. {\bf 53} (1980) 569
\smallskip
\item{14.} T.D. Lee and C.S. Wu, Ann. Rev. Nucl. Sci. {\bf 16} (1966) 511
\smallskip
\item{15.} F. Benatti and R. Floreanini, On the decay law for unstable open
systems, Phys. Lett. B, to appear
\smallskip
\item{16.} R. D\"umcke and H. Spohn, Z. Physik B {\bf 34} (1979) 419
\smallskip
\item{17.} L.J. Garay, Spacetime foam as quantum thermal bath,
{\tt gr-qc/9801024}
\smallskip
\item{18.} J. Ellis, N.E. Mavromatos and D.V. Nanopoulos, Phys. Lett.
{\bf B293} (1992) 142; Int. J. Mod. Phys. {\bf A11} (1996) 1489
\smallskip
\item{19.} V. Gorini and A. Kossakowski, J. Math. Phys. {\bf 17} (1976) 1298
\smallskip
\item{20.} A. Frigerio and V. Gorini, J. Math. Phys. {\bf 17} (1976) 2123
\smallskip
\item{21.} A. Frigerio, C. Novellone and M. Verri, Rep. Math. Phys. 
{\bf 12} (1977) 279
\smallskip
\item{22.} P.F. Palmer, J. Math. Phys. {\bf 18} (1977) 527
\smallskip
\item{23.} W.H. Louisell, {\it Quantum Statistical Properties of Radiation},
(Wiley, New York, 1973)
\smallskip
\item{24.} C.W. Gardiner, {\it Quantum Noise} (Springer, Berlin, 1992)
\smallskip
\item{25.} H.S. Green, Phys. Rev. {\bf 90} (1953) 270
\smallskip
\item{26.} Y. Ohnuki and S. Kamefuchi, {\it Quantum Field Theory and
Parastatistics}, (Springer, Berlin, 1982)
\smallskip
\item{27.} R. Floreanini and L. Vinet, Phys. Rev. D {\bf 44} (1992) 3851 
\smallskip
\item{28.} O.W. Greenberg, Phys. Rev. Lett. {\bf 64} (1990) 705;
Phys. Rev. D {\bf 43} (1991) 4111
\smallskip
\item{29.} A. Strominger and C. Vafa, Phys. Lett. {\bf B379} (1996) 99
\smallskip
\item{30.} C. Callan and J. Maldacena, Nucl. Phys. {\bf B472} (1996) 591
\smallskip
\item{31.} T. Horowitz and A. Strominger, Phys. Rev. Lett. {\bf 77} (1996) 2368
\smallskip
\item{32.} J. Maldacena and A. Strominger, Phys. Rev. D {\bf 55} (1996) 861
\smallskip
\item{33.} For a review, see: J. Maldacena, Black holes in string theory,
PhD-thesis, Princeton 1996, {\tt hep-th/9607235}
\smallskip
\item{34.} T. Banks, W. Fischler, I.R. Klebanov and L. Susskind,
Phys. Rev. Lett. {\bf 80} (1998) 226;
Schwarzchild black holes in Matrix theory II, 1997, {\tt hep-th/9711005}
\smallskip
\item{35.} H. Liu and A.A. Tseytlin, JHEP {\bf 1} (1998) 10
\smallskip
\item{36.} D. Minic, Infinite statistics and black holes in Matrix theory,
Pennsylvania University preprint, 1997, {\tt hep-th/9712202}
\smallskip
\item{37.} For a review, see: D. Bigatti and L. Susskind,
Review of Matrix theory,\hfill\break {\tt hep-th/9712072}
\smallskip
\item{38.} A. Strominger, Phys. Rev. Lett. {\bf 71} (1993) 3397
\smallskip
\item{39.} I.V. Volovich, D-branes, black holes and $SU(\infty)$ gauge theory,
{\tt hep-th/9608137}
\smallskip
\item{40.} For an introduction, see: J. Polchinski, TASI Lectures on D-branes, 
1996, {\tt hep-th/9611050}
\smallskip
\item{41.} T. Altherr and T. Grandou, Nucl. Phys. {\bf B402} (1993) 195
\smallskip
\item{42.} M. Chaichian, R. Gonzalez Felipe and C. Montonen, J. Phys. A
{\bf 26} (1993) 4017
\smallskip
\item{43.} For a discussion, see:
L. Haag, {\it Local Quantum Fields}, (Springer, Berlin, 1996)
\smallskip
\item{44.} H.S. Carslaw, {\it Introduction to the 
Theory of Fourier's series
and Integrals}, (Dover, London, 1930)
\smallskip
\item{45.} E.C. Titchmarsh, {\it Introduction 
to the Theory of Fourier Integrals},
(Clarendon Press, Oxford, 1948)

\bye